\documentstyle[12pt,epsfig]{article}
\def\lsim{\raise0.3ex\hbox{$<$\kern-0.75em\raise-1.1ex\hbox{$\sim$}}}
\def\gsim{\raise0.3ex\hbox{$>$\kern-0.75em\raise-1.1ex\hbox{$\sim$}}}

\def\beq{\begin{equation}}
\def\eeq{\end{equation}}
\def\bea{\begin{eqnarray}}
\def\eea{\end{eqnarray}}
\def\bq{\begin{quote}}
\def\eq{\end{quote}}

\newcommand{\rr}{\mbox{\boldmath $r$}}
\newcommand{\rrn}{\mbox{$r$}}

\def\gappeq{\mathrel{\rlap {\raise.5ex\hbox{$>$}}
{\lower.5ex\hbox{$\sim$}}}}

\def\lappeq{\mathrel{\rlap{\raise.5ex\hbox{$<$}}
{\lower.5ex\hbox{$\sim$}}}}

\def\Toprel#1\over#2{\mathrel{\mathop{#2}\limits^{#1}}}

\begin{document}
\pagestyle{empty}
%\begin{flushright}
%{CERN-TH/2001-265}\\
%hep-ph/th number??\\
%\end{flushright}
%\vspace*{5mm}
\begin{center}
{\bf NUCLEAR HEAVY QUARK PHOTOPRODUCTION IN A SATURATION MODEL }
\\
\vspace*{1cm}
 V.P. Gon\c{c}alves $^{1}$, M.V.T. Machado  $^{1,\,2}$\\
\vspace{0.3cm}
{$^{1}$ Instituto de F\'{\i}sica e Matem\'atica,  Universidade
Federal de Pelotas\\
Caixa Postal 354, CEP 96010-090, Pelotas, RS, Brazil\\
$^{2}$ \rm High Energy Physics Phenomenology Group, GFPAE,  IF-UFRGS \\
Caixa Postal 15051, CEP 91501-970, Porto Alegre, RS, Brazil}\\
\vspace*{1cm}
{\bf ABSTRACT}
\end{center}

\vspace*{1cm} \noindent

\vspace*{1.5cm} \noindent \rule[.1in]{17cm}{.002in}

\vspace{-3.5cm} \setcounter{page}{1} \pagestyle{plain} We
calculate the nuclear inclusive and diffractive cross sections for
heavy quark photoproduction within a phenomenological saturation
approach. The nuclear cross section is obtained by the extension
of the saturation model through Glauber-Gribov formalism. We
predict large  nuclear heavy quark cross sections   at LHC
energies.
  \vspace{1.5cm}

\section{Introduction}

During the last years the study of high  gluon density  effects at
high energy (small $x$) has become  an increasingly active subject
of research, both from experimental and theoretical points of
view. In particular, it has  been observed that the $ep$ deep
inelastic scattering (DIS) data at low $x$ \cite{datasource} can be
successfully described with the help of the saturation model
\cite{golecwus}, which incorporates the main characteristics of
the high density QCD approaches \cite{hdqcd,ayala1,iancu}. In the high energy
regime, the growth of parton distributions should saturate,
possibly forming a color glass condensate \cite{iancu} (For a
pedagogical presentation  see Refs. \cite{mcllec,jamaliancu}),
which is characterized by a bulk momentum scale $Q_s$.
 If the saturation scale is larger than the QCD
scale $\Lambda _{QCD}$, then this regime can be studied using weak
coupling methods.  The magnitude of $Q_s$ is associated to the
behavior of the gluon distribution at high energies, and some
estimates has been obtained. In general, the predictions are
$Q_s\sim 1$ GeV at HERA/RHIC and $Q_s\sim 2-3$ GeV at LHC
\cite{gyusat,vicslope}.

In the phenomenological  analysis presented in Refs.
\cite{golecwus,barkow}, the critical line which marks the
transition to the saturation regime is  smaller than 1 GeV$^2$,
with larger values  predicted only for the next  generation of
$ep$ colliders. However, deep inelastic scattering on nuclei gives
us a new possibility to reach high-density QCD phase without
requiring extremely low values of $x$. The nucleus in this process
serves as an amplifier for nonlinear phenomena expected in QCD at
small $x$, obtaining at the accessible energies at HERA and RHIC
with an $eA$ collider the parton densities which would be probed
only at energies comparable to LHC energies with an $ep$ collider.
In order to understand this expectation and estimate the kinematic
region where the high densities effects should be present, we can
analyze the behavior of the function $\kappa
(x,Q^2) \equiv \frac{3 \pi^2 \alpha_s  }{2 Q^2 } \frac{ xg_A(x,Q^2)}{\pi R^2_A%
} $, which represents the probability of gluon-gluon interaction
inside the parton cascade, and also is  denoted the packing factor
of partons in a parton cascade \cite{GLR,ayala1}. Considering that the condition
$\kappa = 1$ specifies the critical line, which separates between
the linear (low parton density) regime $\kappa \ll 1$ and the high
density regime $\kappa \gg 1$, we can define the saturation momentum scale $%
Q_s$ given by \footnote{No exact definition of the saturation
scale is known so far. In particular, the Eq. (\ref{sat})  is
valid in the double logarithmic approximation. More detailed
analysis imply that the enhancement between the nucleon and
nuclear cases is smaller than that one predicted in this
approximation (See e.g. Refs.  \cite{levtuc,armesto}).}
\begin{eqnarray}
 Q_s^2 (x; A) = \frac{3 \pi^2 \alpha_s  }{2 } \frac{
xg_A(x,Q_s^2(x;A))}{\pi R^2_A}\,\,\,,  \label{sat}
\end{eqnarray}
below which the gluon density reaches its maximum value
(saturates). At any value of $x$ there is a value of $Q^2 =
Q_s^2(x)$ in which the gluonic density reaches a sufficiently high
value that the number of partons stops to rise. This scale depends
on the energy of the process [$xg \propto x^{- \lambda}$ ($\lambda
\approx 0.3$)] and on the atomic number of the colliding nuclei
[$R_A \propto A^{\frac{1}{3}} \rightarrow Q_s^2 \propto
A^{\frac{1}{3}}$], with the saturation scale for nuclear targets
larger than for nucleon ones. Therefore, we expect that the
saturation effects should be manifest in nuclear collisions.
Recently, a phenomenological analysis of the multiplicity
distributions of produced particles in heavy ion collisions at
RHIC, considering the saturation of the nuclear wavefunction,
reproduce the experimental data very well \cite{kharlev}. However,
the current situation is not unambiguous, since other approaches
which do not assume saturation also describe the same set of data.
This result motivates more extensive studies of nuclear collisions
and, in particular, of electron-nucleus collisions at high
energies, where the number of other medium effects is reduced in
comparison with $AA$ collisions. Recently, some authors
\cite{armesto,armsal,teaney} have addressed this subject with
particular emphasis in the behavior of the nuclear structure
functions (For previous studies see e.g. Refs.
\cite{ayala1,qiuesk,ina,vicgayprc}), obtaining predictions which
agree with the scarce experimental data. Since these models rely
on different assumptions, more accurate experimental results for
the nuclear structure function in a large kinematical region are
necessary to probe the high density dynamics. However, these
results compel us to suggest further measurements in
electron-nuclei interactions. 

Our goal in this paper is to investigate the high energy heavy
quark photoproduction on nuclei targets using the saturation
hypothesis. The nuclear photoproduction has been recently studied
in Ref. \cite{barlev}, obtaining a quite reasonable description of
the experimental data,  but the specific topic of heavy quark
production was not  addressed  in that analysis. It is important
to emphasize that for heavy quark production the predictions  are
not dependent of the hypothesis  for the soft region (e.g. the
values of the light quark  masses). Here we study the nuclear
photoproduction of heavy quarks considering the 
 approach proposed in Ref. \cite{armesto}, which is simpler, gives an equally reasonable
 agreement with nuclear data as  other approaches, and is associated with a model which
gives a good description of inclusive and diffractive $ep$
experimental data. A detailed comparison between the distinct
saturation approaches for the nuclear case will be presented in a
separated publication \cite{mvfuturo}.

This paper is organized as  follows. In the next section we
present a brief review of the saturation model  for the nucleon
and nuclear case, and the dipole nuclear cross section is
analyzed. Section \ref{photo} presents our predictions for the
energy and atomic number dependence of the heavy quark
photoproduction cross section, as well as a detailed analysis of
the overlap functions, which allow us to verify what is mean
dipole size that contributes for this process. As a by product, in
Section \ref{dif} the diffractive production of heavy quarks in
photonuclear collisions is also considered in the saturation
model. Finally, in Section \ref{conc} we summarize our
conclusions.

\section{Nuclear cross sections in a saturation model}

In this section we shortly  review the main concepts and formulae
of the saturation model which is based on the  color dipole
approach. The latter gives  a simple unified picture of inclusive
and diffractive processes which provides a large phenomenology on
DIS regime. In this approach,  the scattering process can be seen
in the target rest frame as a succession in time of three
factorizable subprocesses: i) the photon fluctuates in a
quark-antiquark pair, ii) this color dipole interacts with the
target and, iii) the quark pair annihilates in a virtual photon.
Using as kinematic variables the $\gamma^* N$ c.m.s. energy
squared $s=W^2=(p+q)^2$, where $p$ and $q$ are the target and the
photon momenta, respectively, the photon virtuality squared
$Q^2=-q^2$ and the Bjorken variable $x=Q^2/(W^2+Q^2)$, the
corresponding total cross section  reads as \cite{nik,golecwus}
\begin{eqnarray}
\sigma_{T,L} (x,Q^2)  = \int_0^1
dz\, \int d^2\rr \, |\Psi_{T,\,L} (z,\,\rr,\,Q^2)|^2 \, \sigma_{dip}^{\mathrm{target}}
(\tilde{x},\,\rr^2)\,,
\label{sigmatot}
\end{eqnarray}
where
\begin{eqnarray}
 |\Psi_{T} (z,\rr,\,Q^2)|^2 & = &  \frac{6\alpha_{\mathrm{em}}}{4\,\pi^2} \,
 \sum_f e_f^2 \, \left\{[z^2 + (1-z)^2]\, \varepsilon^2 \,K_1^2(\varepsilon \,\rrn)
 + m_f^2 \, \,K_0^2(\varepsilon\, \rrn)
 \right\}\,,\label{wtrans}\\
 |\Psi_{L} (z,\rr,\,Q^2)|^2 & = & \frac{6\alpha_{\mathrm{em}}}{\pi^2} \,
\sum_f e_f^2 \, \left\{Q^2 \,z^2 (1-z)^2 \,K_0^2(\varepsilon\, \rrn)
\right\}\,,
\label{wlongs}
 \end{eqnarray}
 are the squared photon wave
function for transverse ($T$) and longitudinal ($L$) photons,
respectively. The variable $\rr$ defines the relative transverse
separation of the pair (dipole) and $z$ $(1-z)$ is the
longitudinal momentum fractions of the quark (antiquark). The
auxiliary variable $\varepsilon^2=z(1-z)\,Q^2 + m^2_f$ depends on
the quark mass, $m_f$. The $K_{0,1}$ are the McDonald functions
and the summation is performed over the quark flavors.

For electron-proton interactions, the dipole  cross section
$\sigma_{dip}^{\mathrm{p}}$, describing the dipole-proton
interaction is substantially affected by non-perturbative content.
There are several phenomenological implementations for this
quantity \cite{dipolos}. The main feature of these approaches is  to be
able to match the soft (low $Q^2$) and hard (large $Q^2$) regimes
in an unified way. In the present work we follow the quite
successful saturation model \cite{golecwus}, which
interpolates between the small and large dipole configurations,
providing color transparency behavior, $\sigma_{dip}\sim \rr^2$,
as $\rr \rightarrow 0$ and constant behavior, $\sigma_{dip}\sim
\sigma_0$, at large dipole separations. The parameters of the
model have been obtained from an adjustment to small $x$ HERA
data. Its parameter-free application to diffractive DIS has been
also quite successful \cite{golecwus} as well as its
extension to virtual Compton scattering \cite{Favart_Machado},
vector meson production \cite{Caldwell_Mara} and two-photon
collisions \cite{Kwien_Motyka}.   The parameterization for the
dipole cross section takes the eikonal-like form,
\begin{eqnarray}
\sigma_{dip}^{\mathrm{p}} (\tilde{x}, \,\rr^2) & = & \sigma_0 \,
\left[\, 1- \exp \left(-\frac{\,Q_s^2(x)\,\rr^2}{4} \right) \, \right]\,,
\label{gbwdip}\\ Q_s^2(x) & = & \left( \frac{x_0}{\tilde{x}}
\right)^{\lambda} \,\,\mathrm{GeV}^2\,,
\end{eqnarray}
where the saturation scale $Q_s^2$ defines the onset of the
saturation phenomenon, which depends on energy. The parameters
were obtained from a fit to the HERA data producing
$\sigma_0=23.03 \,(29.12)$ mb, $\lambda= 0.288 \, (0.277)$ and
$x_0=3.04 \cdot 10^{-4} \, (0.41 \cdot 10^{-4})$ for a 3-flavor
(4-flavor) analysis~\cite{golecwus} (See Refs.
\cite{barkow,teaney} for improvements of this model). An
additional parameter is the effective light quark mass, $m_f=0.14$
GeV, consistent with the pion mass. It should be noticed that the
quark mass plays the role of a regulator for the photoproduction
($Q^2=0$) cross section. The light  quark mass is one of the
non-perturbative inputs in the model. The charm quark mass is
considered to be $m_c=1.5$ GeV. A smooth transition to the
photoproduction limit is obtained with a modification of the
Bjorken variable as,
\begin{eqnarray}
\tilde{x}= x\, \left( \, 1+ \frac{4\,m_f^2}{Q^2}
\,\right)=\frac{Q^2 + 4\,m_f^2}{W^2} \,.
\end{eqnarray}
The
saturation model is suitable in the region below $x=0.01$ and the
large $x$ limit needs still a consistent  treatment. Making use of
the dimensional-cutting rules, here we supplement
 the dipole cross section, Eq. (\ref{gbwdip}), with a threshold factor
$(1-x)^{n_{\mathrm{thres}}}$, taking $n_{\mathrm{thres}}=5$ for a
3-flavor analysis and $n_{\mathrm{thres}}=7$ for a 4-flavor one.
This procedure ensures consistent description of heavy quark
production at the fixed target data \cite{Mariotto_Machado}.

Before going further, some comments on the impact parameter  dependence of  the saturation model are concerned. The implicit assumption in the approach is  that the proton is  treated as being homogeneous in the transverse plane. In such case, the impact parameter profile is given by the Heaviside function, $S(b)=\Theta\,(b_0-b)$. The profile is considered to be peacked at central impact parameter, namely  at $b=0$. Actually, this procedure is oversimplified and more realistic profiles can be considered. For phenomenological purposes a gaussian or a hard sphere assumption are commonly taken into account. Recently, the impact parameter dipole saturation model \cite{teaney} was developed, recovering the known Glauber-Mueller dipole cross section. There, distinct shapes for  $S(b)$ were considered and their parameters constrained from data on  $t$-dependence of the $J/\Psi$ photoproduction.  

Let us discuss the extension of the saturation  model for the
photon-nucleus interactions. Here, we follow the simple procedure
proposed in Ref. \cite{armesto}, which consists of an extension to
nuclei, using the Glauber-Gribov picture \cite{gribov}, of the
saturation model discussed above, without any new parameter (For 
similar approaches see e.g.  Refs. \cite{ayala1,nik2}). There, the nuclear
version is obtained replacing the dipole-nucleon cross section in
Eq. (\ref{sigmatot}) by the nuclear one,
\begin{eqnarray}
\sigma_{dip}^{\mathrm{A}} (\tilde{x}, \,\rr^2, A)  = \int d^2b \,\, 2
\left\{\, 1- \exp \left[-\frac{1}{2}\,A\,T_A(b)\,\sigma_{dip}^{\mathrm{p}} (\tilde{x}, \,\rr^2)  \right] \, \right\}\,,
\label{sigmanuc}
\end{eqnarray}
where $b$ is the impact parameter of the center of the dipole
relative to the center of the nucleus and the integrand gives the
total dipole-nucleus cross section for a  fixed impact parameter.
The nuclear profile function is labelled by $T_A(b)$, which will
be obtained from a 3-parameter Fermi distribution for the nuclear
density \cite{devries}. The above equation sums up all the multiple elastic rescattering diagrams of the $q \overline{q}$ pair
and is justified for large coherence length, where the transverse separation $r$ of partons in the multiparton Fock state of the photon becomes as good a conserved quantity as the angular momentum, {\it i. e.} the size of the pair $r$ becomes eigenvalue
of the scattering matrix. It is important to emphasize that for very small values of $x$, other diagrams beyond the multiple Pomeron exchange considered here should contribute ({\it e.g.} Pomeron loops) and a more general approach for the high density (saturation) regime must be considered. However, we believe that this approach allows us to obtain lower limits of the high density effects in the RHIC and LHC kinematic range. Therefore, at first glance,  the region of applicability of
this  model should be at  small values of  $x$, i.e. large
coherence length, and for not too high  values of virtualities,
where the implementation of the DGLAP evolution should be
required. Therefore, the approach is quite suitable for the
analysis of heavy quark photoproduction in the  RHIC and LHC
kinematical ranges.

\begin{figure}[t]
\begin{tabular}{cc}
\psfig{file=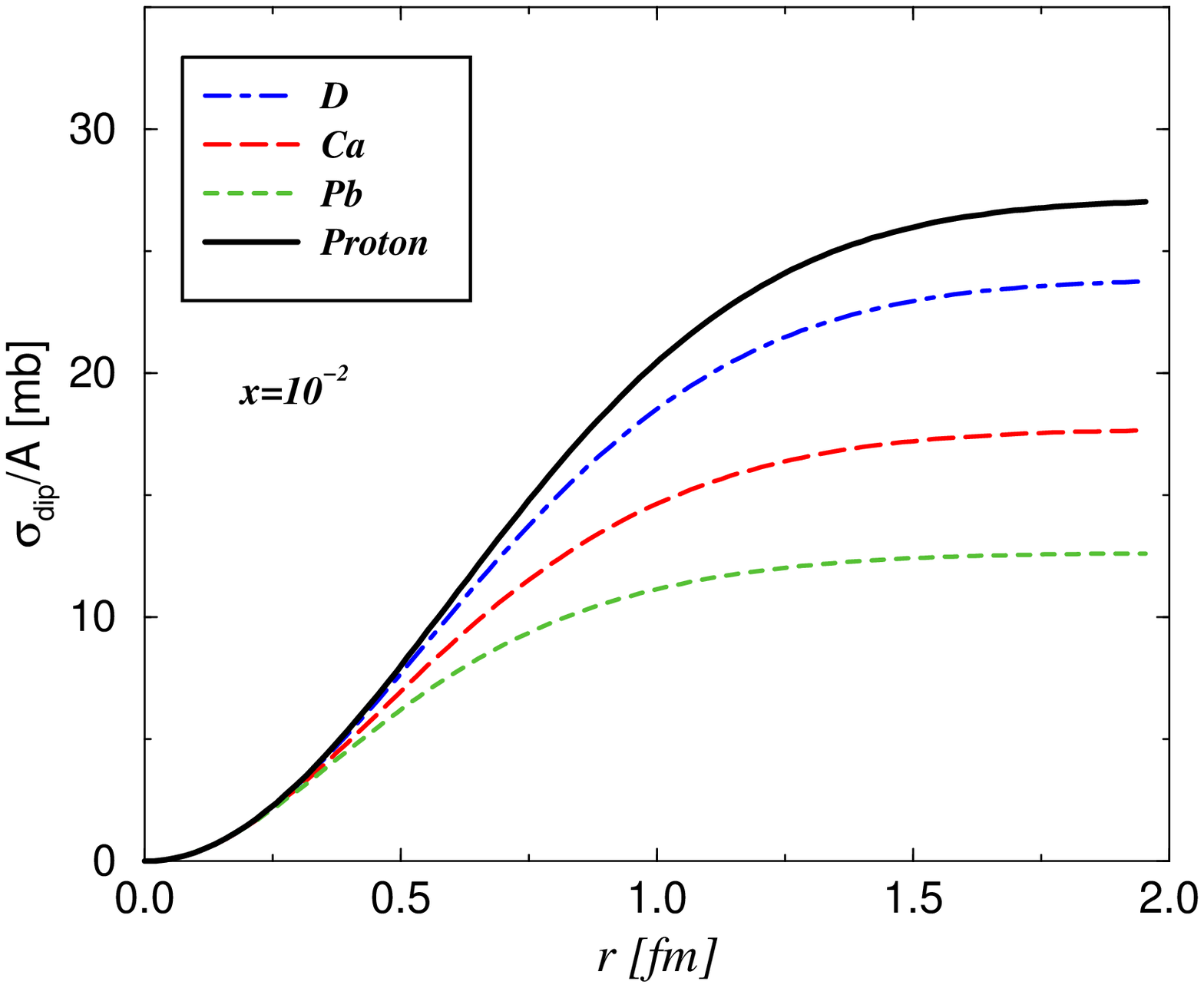,width=70mm} & \psfig{file=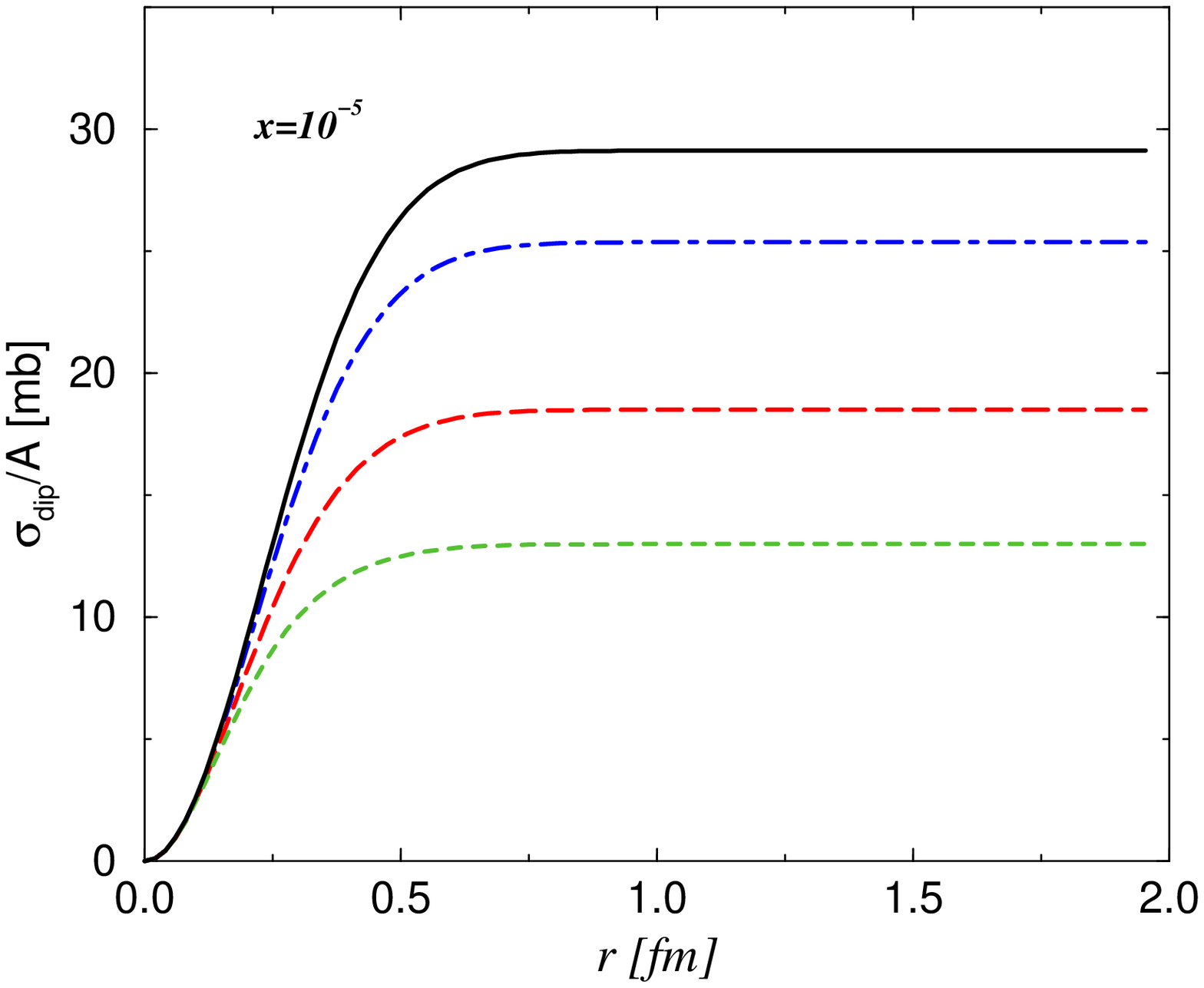,width=70mm}
\end{tabular}
 \caption{\it The nuclear dipole cross section as a function of
 dipole size $r$ for different $x$ values and distinct $A$, including the proton case.
  The results are normalized to $A$.}
\label{fig1}
\end{figure}

In order to investigate the dependence of the nuclear dipole cross
section on the  dipole size, in Fig. \ref{fig1} are shown our
results for $\sigma_{dip}^{\mathrm{A}}$  as a function of the
dipole size  at two fixed values of $x$. We have select different
nuclei $A=$ D, Ca and Pb as well the proton case at $x=10^{-2}$
and $x=10^{-5}$, with the dipole cross section normalized to $A$.
The behavior on $r$ is the same as the proton case,  saturating at
large  values of $r$.  At small dipole sizes the result is almost
independent of $A$, whereas at large dipole sizes the nuclear
suppression is quite substantial and dependent on $A$. For
instance, the reduction in the overall normalization reaches to a
factor about 3 for lead in comparison with the nucleon case at
small $x$.  This result has sizeable consequences on processes
where the large dipole contribution is important, for example in
the transversal component of  $F_2^A$ and diffractive production
of light quarks. For heavy quarks, the situation is different
since the corresponding wave functions selects mostly small
dipoles configurations, as we will show later on. Having addressed
the main concepts and definitions, in the next section one
presents our results for the heavy quark production within the
color dipole picture supplemented by the saturation model
discussed above. Here, we will focus on photoproduction of charm
and bottom in the kinematical range relevant to RHIC and LHC.

\section{Nuclear heavy quark photoproduction}
\label{photo}

In this section  we compute the nuclear cross section for the
photoproduction of charm and bottom quarks. In this case the
nonperturbative input associated with the light quark mass is
avoided and we are left with the hard scale given by the heavy
quark masses. Here, the following values are taken into account,
$m_c=1.5$ GeV and $m_b=4.5$ GeV, to be consistent with the
previous analysis using the saturation model in the nucleon case.
Notice that the use of a different choice implies  a distinct
overall normalization in the final result.

The cross section for heavy quark photoproduction on nuclei
targets  is given by
\begin{eqnarray}
\sigma_{tot}^{\gamma\,A} (W, A)  = \int_0^1
dz\, \int d^2\rr \, |\Psi_{T} (z,\,\rr,\,Q^2=0)|^2 \, \sigma_{dip}^{\mathrm{A}}
(\tilde{x},\,\rr^2,A)\,,
\label{sigmaphotd}
\end{eqnarray}
where the longitudinal  contribution is suppressed [See Eq.
(\ref{wlongs}) at $Q^2=0$] and $e^2_{c,\,b}= 4/9$ and $1/9$,
respectively. Accordingly, $m_f=m_{c,\,b}$ in Eq. (\ref{wtrans})
and the parameters for the dipole cross section are taken from the
4-flavor analysis.
The formula (\ref{sigmaphotd}) sums up in a compact form all the elastic + inelastic rescattering diagrams of the heavy quark 
pair with the nucleus.

\begin{figure}[t]
\begin{tabular}{cc}
\psfig{file=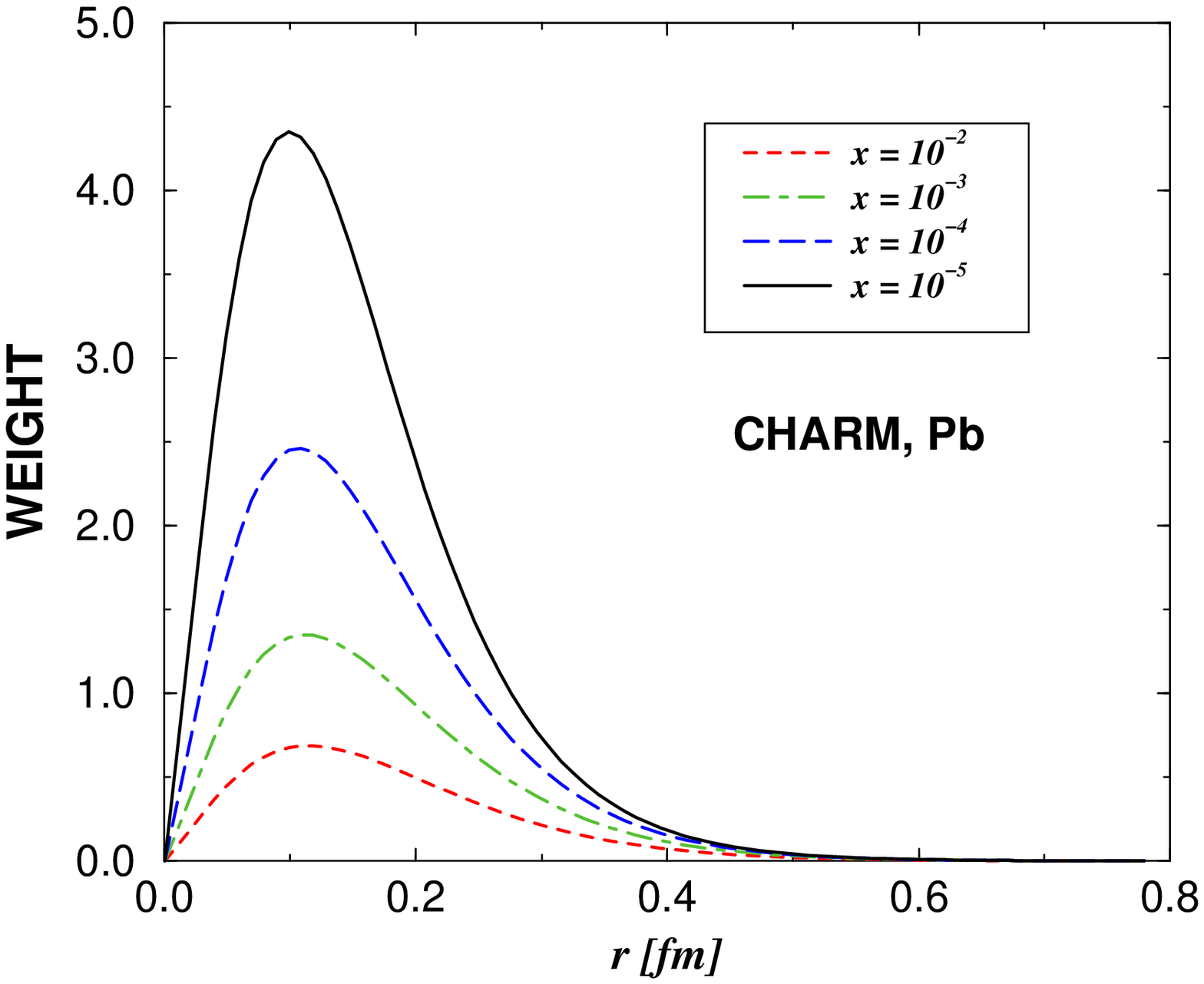,width=70mm} & \psfig{file=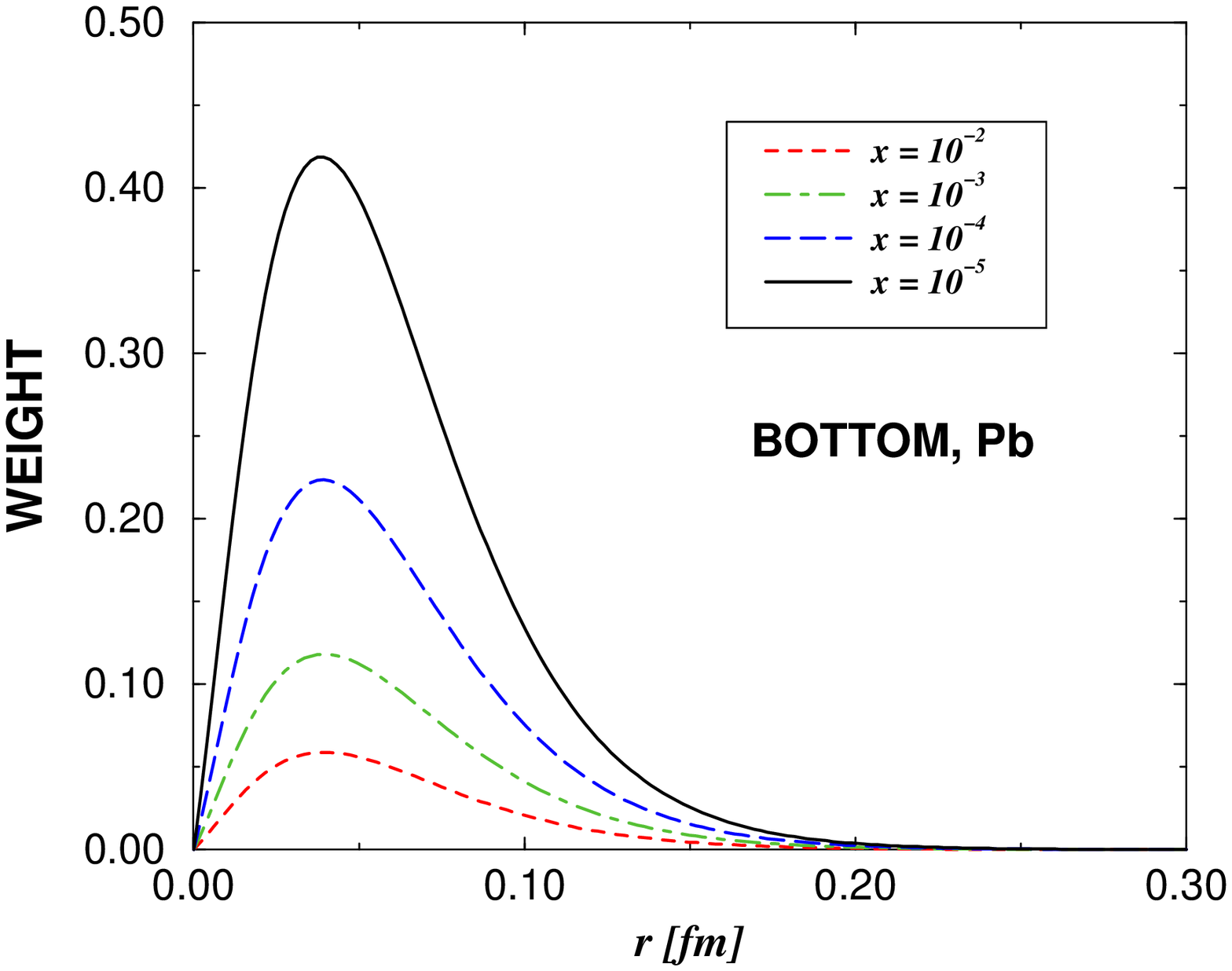,width=72mm}
\end{tabular}
\caption{\it The $r$ dependence of the  photon-nucleus overlap function
for charm and bottom production,  different $x$ and
$A=Pb$.}
\label{fig2}
\end{figure}

Before  presenting our  results for energy dependence of the cross
section, we can investigate  the mean dipole size dominating the
nuclear heavy quark  photoproduction. We define the photon-nucleus
overlap function which reads as,
\begin{eqnarray}
{\cal W} \,(\tilde{x},\rr,A)  =  2\pi \rrn \, \int dz\, |\Psi_{T}
(z,\,\rr)|^2 \, \sigma_{dip}^{\mathrm{A}} (\tilde{x},\,\rr^2,A)\,.
\label{overlap}
\end{eqnarray}
In Fig. \ref{fig2} are  shown the overlap function for the charm
and bottom production as a function of dipole size. They are
computed for lead nucleus and for different $x$. In the charm
case, the distribution is peaked at approximately $r=0.1$ fm,
whereas for the bottom case this value is shifted to $r=0.05$ fm, which agree with the theoretical expectation that the  $q\overline{q}$ pairs have a typical transverse size $\approx \, 1/m_f$ \cite{nik2}.
Therefore, the main contribution to the cross section comes from
the small  dipole sizes, i. e. from the perturbative regime. In
contrast, for light quarks a broader $r$ distribution is obtained,
peaked for large values of the pair separation, implying that
nonperturbative contributions cannot be disregarded in that case.
For sake of illustration, the transition region between
perturbative and nonperturbative regimes ranges at $Q^2\sim 1/r^2
\simeq 1$ GeV$^2$, which means perturbative domain for $r\leq
0.2-0.4$ fm. Concerning the high density effects, its value is more
sizeable at large dipole configurations (See Fig. \ref{fig1}),
meaning that for heavy quark photoproduction we should expect that
the associated modifications will be small.

\begin{figure}[t]
\centerline{\psfig{file=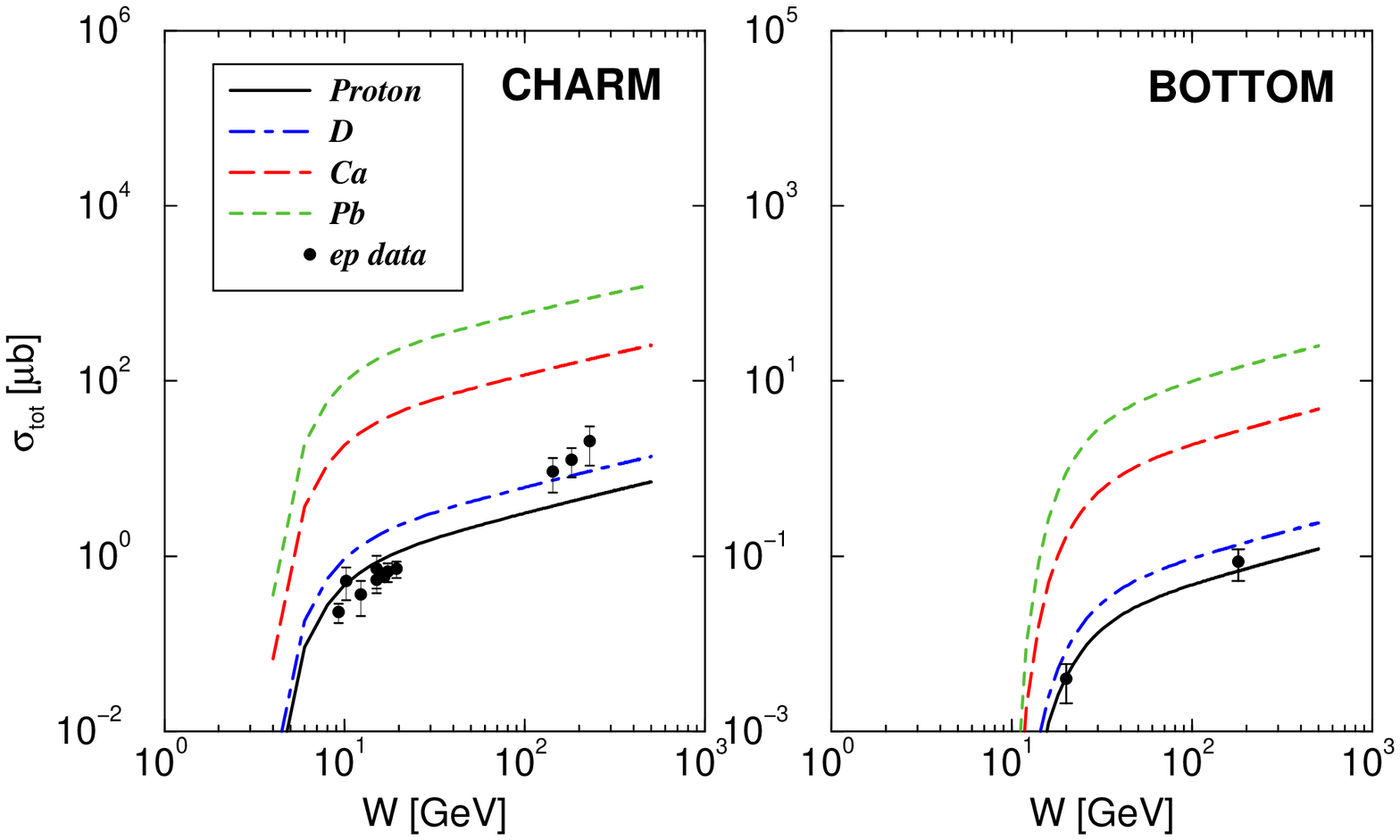,width=120mm}}
 \caption{\it The total nuclear cross section for charm and bottom photoproduction as
 a function of cms energy $W$ and distinct nuclei.}
 \label{fig3}
\end{figure}

In Fig. \ref{fig3} are shown the results for the charm and bottom
photoproduction cross section as a function of energy for
different nuclei, including the proton case. The results present
mild growth on $W$ at high energies stemming from the saturation
model, whereas the low energy region is consistently described
through the threshold factor. For the proton, the experimental data from HERA \cite{datahera} and fixed target collisions 
\cite{fixedtarget}  are also included for sake of comparison. The result for charm underestimates data by a factor 2 at $W_{\gamma p} \simeq 200$  GeV, whereas is consistent with the measurements of bottom cross section. Concerning charm production, the measured cross sections present a well known steeper behavior on energy even at electroproduction, suggesting that further resummations in the original saturation model are needed in order to produce the  larger growth on energy  appearing in the charm measurements. We believe that the better  result for bottom happens to be  a mismatch between a large uncertainty in the experimental measurement and the lower bottom mass $m_b=4.5$ GeV  considered here.   For the nuclear case,  we predict that their absolute values are rather large, reaching $\approx 2 \cdot 10^3$  and $\approx 40 \, \mu b$
for charm and bottom for lead at $W = 10^3$ GeV. It is important
to emphasize that our results agree with the predictions obtained for charm photoproduction 
in the Ref. \cite{braunarm}, where the heavy flavor
leptoproduction off the nucleus has been addressed considering an
approach which resums the fan diagrams of BFKL pomerons, with
initial condition for the evolution at $x = 0.01$ given by the Eq.
(\ref{gbwdip}). For bottom photoproduction, our results  are smaller than Ref. \cite{braunarm}. We believe that our results are  reliable, since they are consistent with the simple expectation   $\sigma_{\gamma A \rightarrow b\overline{b} X} \approx A \times \sigma_{\gamma p \rightarrow b\overline{b} X}$ for the color transparency regime.  A check of the saturation model for the proton
case can be found in Ref. \cite{Mariotto_Machado} and a comparison
with other approaches in Ref. \cite{mvfuturo}. As expected from
the discussion in the previous paragraph,   the nuclear cross
sections can be reasonably approximated by $A$ times the nucleon
cross section (Color transparency) \cite{brodmue}, since the process is dominated by the scattering of
small size dipoles. This result is consistent with the behavior of
the dipole cross section presented in Fig. \ref{fig1}, where we
have seen that sizeable nuclear effects are only important at
large dipole sizes.

\begin{figure}[t]
\begin{tabular}{cc}
\psfig{file=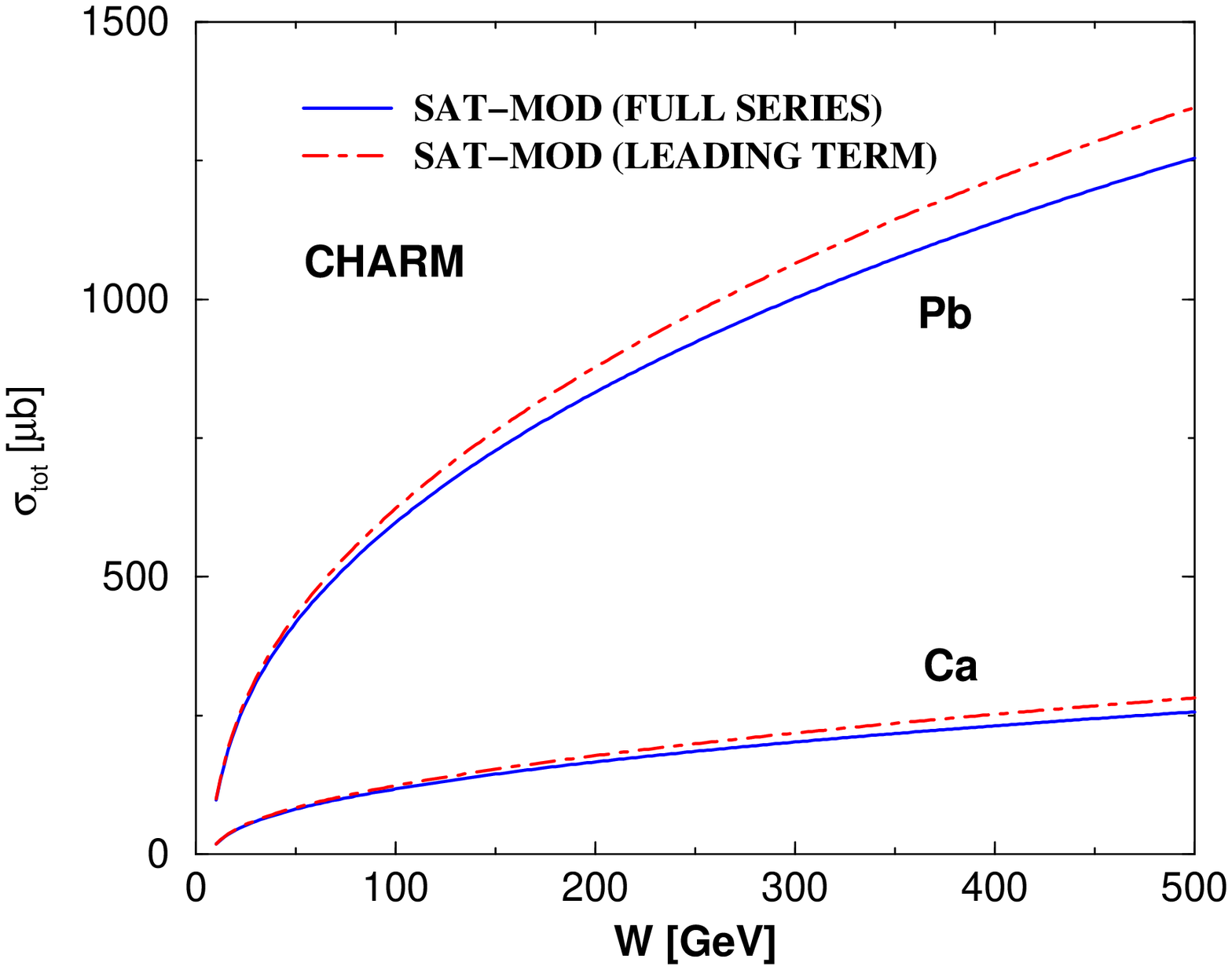,width=79mm} &
\psfig{file=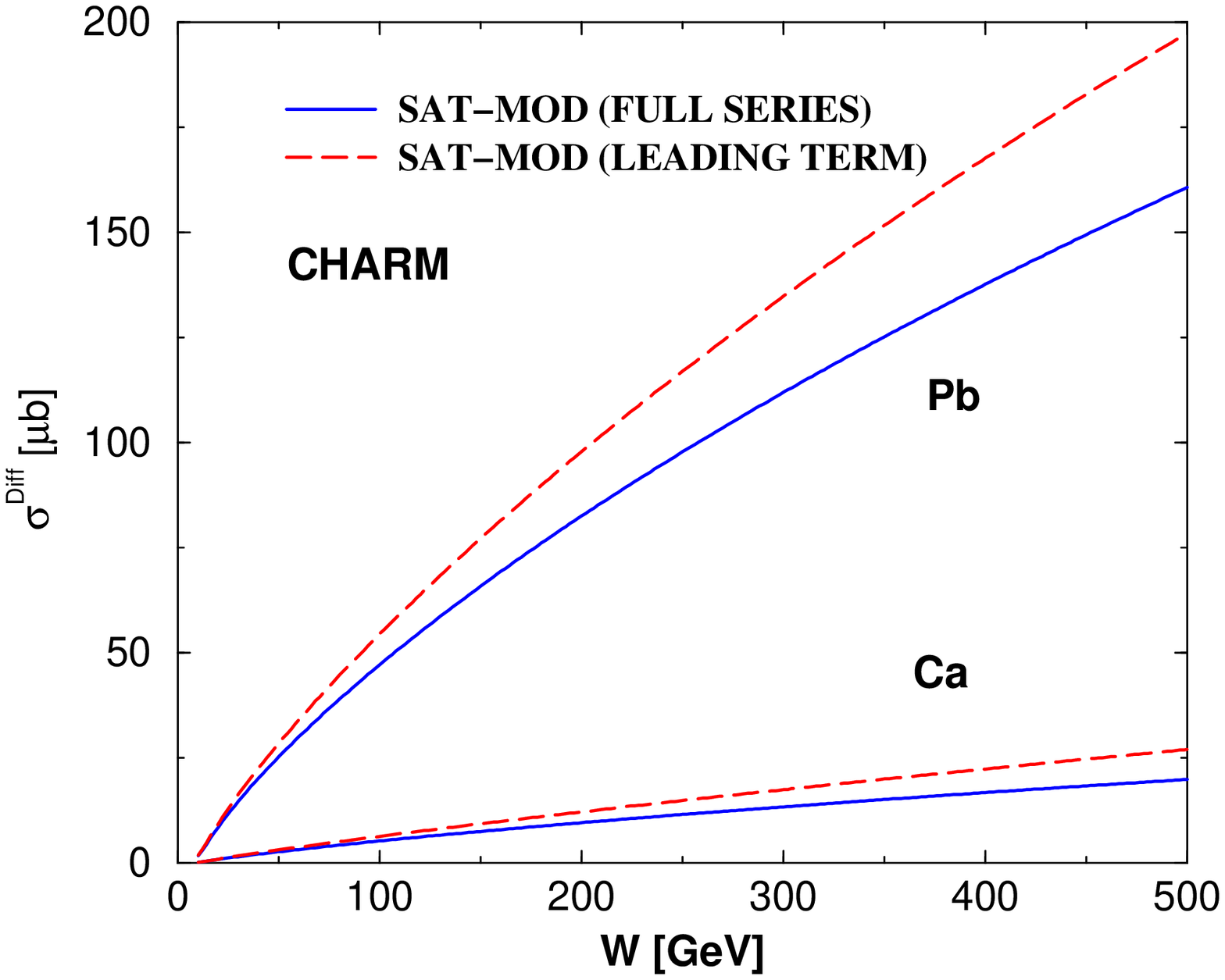,width=77mm}\\
(a) & (b)
\end{tabular}
 \caption{\it The result for the  (a) inclusive and (b) diffractive  nuclear charm photoproduction cross
section as a function of c.m.s energy for calcium and lead. The
solid line corresponds to the full dipole nucleon cross section of
the saturation model (full series) and the dot-dashed line is
obtained using only the leading term. }
 \label{fig4}
\end{figure}

Here, an additional issue should be  addressed.  The
dipole-nucleon cross section  resums higher twist contributions
and also ensures the unitarity requirements, producing a constant
value at large $r$. In order to investigate the effects of this
resummation in the dipole-nucleus cross section, we show in Fig.
\ref{fig4} (a)  our results for the nuclear photoproduction cross
section considering as input the full dipole-nucleon cross section
(solid lines), Eq. (\ref{gbwdip}), and  only the leading term in
the expansion of $\sigma_{dip}^{\mathrm{p}}$ (dot-dashed lines).
They were computed for charm quark and considering the calcium and
lead nuclei. The effect is almost negligible for light nucleus,
whereas it is more sizeable for heavy ones at high energies. This
result demonstrate that  nuclear heavy quark production is not
sensitive to the assumption of a saturated nucleon. In contrast,
the behavior of the  nuclear structure function depends on the
saturation effects at the nucleon level, as shown in Ref.
\cite{vicgayprc}.

\section{Diffractive photoproduction of heavy quarks}
\label{dif}

Let us now to compute the diffrative production of heavy  quarks.
This process was analyzed for $ep$ collisions in Refs.
\cite{nikopen,MartinLevin,Diehl}.
 In terms of the
$S$ matrix at a given impact parameter of the collision, $S(b)$,
the total and elastic cross sections for the deep inelastic
scattering on a nucleus are given by \cite{muellerepja}
\begin{eqnarray}
\sigma_{tot} = 2 \int d^2b \,[1 -S(b)] \,\,,\\
\sigma_{el} = \int d^2 b \, [1 -S(b)]^2\,\,, \label{mats}
\end{eqnarray}
which demonstrate how easily the elastic and total cross sections
could be obtained from one another. Identifying  $1 - S(b)$ with
the expression within the brackets in Eq. (\ref{sigmanuc}) and
considering that diffraction of the photon on the target can be
thought  as an elastic scattering of each dipole off the nucleus,
we obtain that the nuclear diffractive cross section is
\begin{eqnarray}
\sigma_{T,L}^{\mathrm{Diff}} (x,Q^2)  = \int d^2b \,\int_0^1 dz\,
\int d^2\rr \, |\Psi_{T,\,L} (z,\,\rr,\,Q^2)|^2 \, \left\{1- \exp
\left[-\frac{1}{2}\,A\,T_A(b)\,\sigma_{dip}^{\mathrm{p}}
(\tilde{x}, \,\rr^2)  \right] \right\}^2\,. \label{sigmadif}
\end{eqnarray}
For heavy quark photoproduction we have that only the transverse
polarization contributes and the wave function is dependent on the
heavy quark masses.

\begin{figure}[t]
\begin{tabular}{cc}
\psfig{file=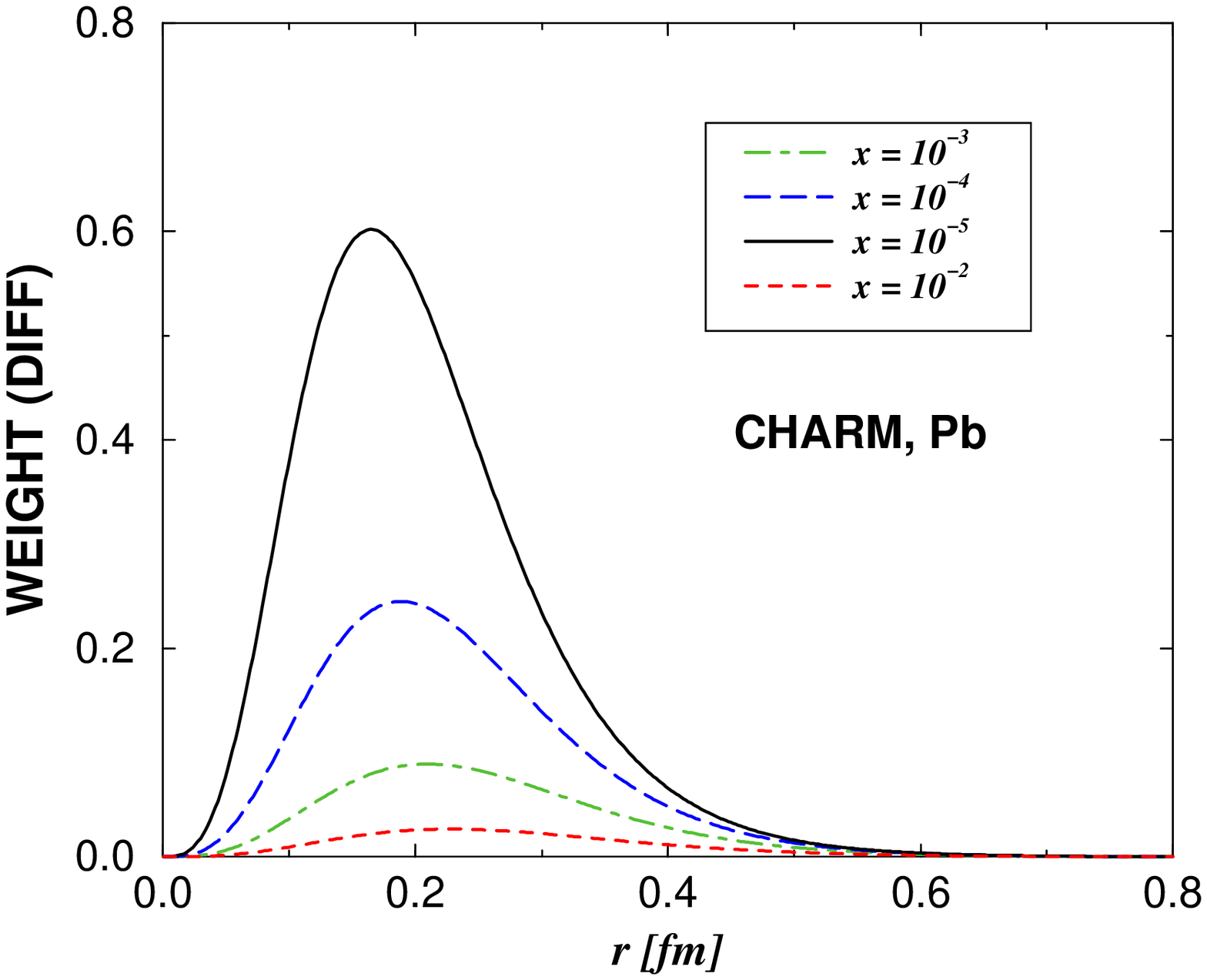,width=75mm} &
\psfig{file=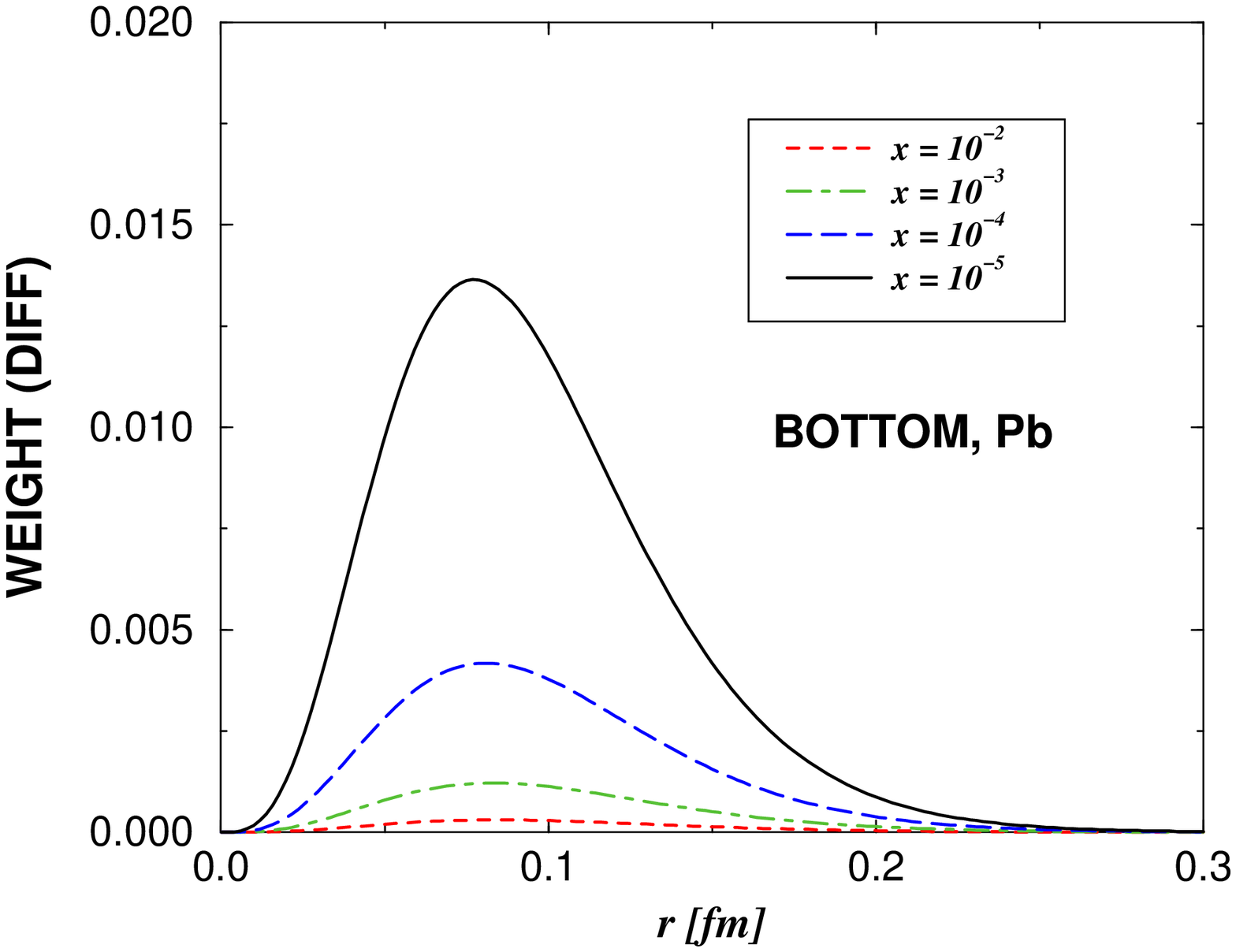,width=79mm}
\end{tabular}
\caption{\it The $r$ dependence of the diffractive photon-nucleus
overlap function for charm and bottom production, different $x$
and $A=Pb$. } \label{fig5}
\end{figure}

Similar analysis of mean size dipole dominance can be made for the
diffractive case. Introducing the  the diffractive overlap
function defined by
\begin{eqnarray}
{\cal W}^{\mathrm{Diff}} \,(\tilde{x},\rr,A)  =  2\pi \rrn \, \int
d^2b \, \int dz\, |\Psi_{T} (z,\,\rr)|^2 \, \left\{1- \exp
\left[-\frac{1}{2}\,A\,T_A(b)\,\sigma_{dip}^{\mathrm{p}}
(\tilde{x}, \,\rr^2)  \right] \right\}^2\,, \label{overlapdif}
\end{eqnarray}
in Fig. \ref{fig5} we present our results for the $r$ dependence
of this function for different values of $x$ and $A=Pb$. We can
see that the distribution is broader and its  maximum  is shifted to  large values of $r$ in comparison with
the inclusive case. This result is expected since the diffractive production has a large amount of 
nonperturbative contributions. However, similarly to the inclusive case, the mean size dipole occur in the 
perturbative regime, with the large distance still strongly suppressed. 

\begin{figure}[t]
\centerline{\psfig{file=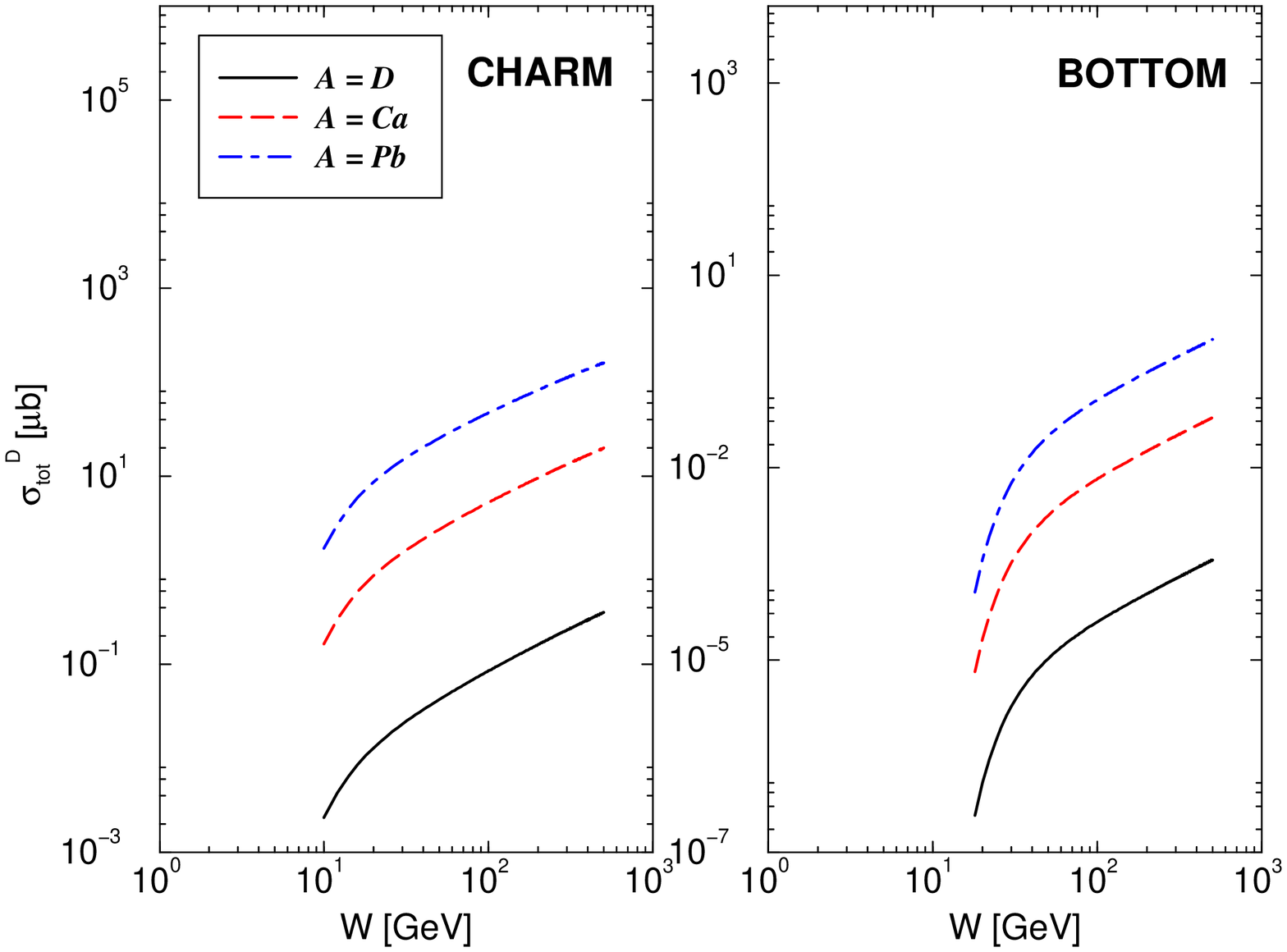,width=120mm}} 
\caption{\it The diffractive nuclear cross section for charm and  bottom
photoproduction as a function of c.m.s. energy $W$ and for
distinct nuclei.} \label{fig6}
\end{figure}

In Fig. \ref{fig6} the results are shown for the charm and
bottom diffractive cross section as a function of energy for
different nuclei. The values at high energies are sizeable, reaching to   
250 $\mu b$ and 0.8 $\mu b$ for charm and bottom production, considering  
lead at $W=10^{3}$ GeV. These values are in agreement with a 10\% contribution from diffractive scattering to the
 total cross section observed  in photon-nucleon reactions. Concerning the dependence on $A$,
 the results present an approximate  behavior $\sigma^{Diff} \simeq A^{4/3}$, as expected for the interaction of small dipoles with the nucleus \cite{nzz,HERAeA}. We also have checked
 the diffractive production of heavy quarks for the proton case, with a cross section 
 of order 100 nb at $W=200 $ GeV for diffractive open charm production, being consistent with
 previous theoretical estimates \cite{Diehl}. There, it was found a value of 60 nb at the 
same energy in the photoproduction case. A more detailed analysis for the proton case should be addressed elsewhere.

Finally, in Fig. \ref{fig4} (b) we analyze the dependence of the diffractive nuclear cross section in considering or not a saturated proton, namely saturation effects in the nucleon level.  We can see that while for a light nuclei the cross section is not sensitive to the assumption of a saturated nucleon, for a heavy nuclei the result differs by  about $20 \%$. Therefore, the analysis of the diffractive cross section can be useful to constraint the nucleon dynamics.

\section{Summary and Conclusions}
\label{conc}

In this paper we have calculated the nuclear inclusive and
diffractive cross sections for heavy quark photoproduction within
 a phenomenological saturation model. In such  model the nuclear cross section is
obtained  through the Glauber-Gribov formalism. Since it describes
reasonably the experimental data for the nuclear structure
function, we are confident in extending this model for the nuclear
heavy quark photoproduction case. Moreover, it is simple and
relies on  a model which gives a good description of inclusive and
diffractive $ep$ experimental data. This model should be valid
until the full non-linear  evolution effects become  important,
which implies the consideration of the Pomeron loops beyond the
multiple scattering on single nucleons estimated in the present
framework. We have verified that the main contribution of the high
density effects occur for large pair separation, while for dipoles
of small size they are almost negligible. We have investigated the
mean dipole size dominance and have verified that the heavy quark
cross section is dominated by small dipole configuration, in
contrast to the light quark case. Consequently, we have that the
nuclear heavy quark production is not strongly modified by the
high density effects. We predict absolute values for the cross
section rather large, being about 2 mb and 0.04 mb for charm and
bottom, respectively, for lead and $W = 1$ TeV. These values are
similar to those resulting from the resummation of the fan
diagrams of BFKL pomerons. Furthermore, we have computed the
diffractive nuclear photoproduction of heavy quarks. This quantity is 
more sensitive to the nonperturbative sector, corresponding to a larger dipole sizes. This means
 that the cross section is dominated by larger dipole configurations than in the inclusive case.  The results 
take values about  250 $\mu b$ and 0.8 $\mu b$ for charm and bottom for lead at  $W = 1$ TeV. Concerning the 
$A$-dependence, we have found a behavior proportional to $A$ and $A^{4/3}$ for the inclusive and diffractive cross sections, respectively,  in agreement with  
theoretical expectations associated with  the color transparency regime.

\section*{Acknowledgments}
 M.V.T.M. thanks the support of the High Energy Physics Phenomenology Group at the Institute of Physics, GFPAE IF-UFRGS, Porto Alegre. This work was partially financed by the Brazilian funding agencies CNPq and FAPERGS.

\end{document}